\documentclass[12pt]{iopart}
\usepackage{txfonts}
\usepackage{graphicx}
\usepackage{dcolumn}
\usepackage{bm}
\usepackage{hyperref}

\begin{document}

\title{Network growth approach to macroevolution}

\author{Shao-Meng Qin, Yong Chen\footnote[2]{To whom correspondence should be addressed. Email: \tt{ychen@lzu.edu.cn}}, and Pan Zhang}

\address{Institute of Theoretical Physics, Lanzhou University, Lanzhou $730000$, China}

\date{\today}

\begin{abstract}
We propose a novel network growth model coupled with the competition interaction to
simulate macroevolution. Our work shows that the competition plays an important
role in macroevolution and it is more rational to describe the interaction
between species by network structures. Our model presents a complete picture of the
development of phyla and the splitting process. It is found that periodic mass
extinction occurred in our networks without any extraterrestrial factors and the
lifetime distribution of species is very close to fossil record. We also perturb
networks with two scenarios of mass extinctions on different hierarchic levels in
order to study their recovery.
\end{abstract}

\pacs{87.23.-n, 89.75.Hc, 07.05.-t}

\submitto{\NJP}
\maketitle

\section{\label{sec1}INTRODUCTION}

The ecosystem, which is formed from a myriad of interactions between various
species, is one of the best-known examples of complexity. During the last decade,
theoretical research on coevolution of species and the statistics of extinctions
have been strongly influenced by the pioneering interdisciplinary work of Per Bak
and his collaborators~\cite{pbak1,pbak2,pbak3,pbak4}. Many simple and delicate
models have been able to explain a large number of phenomena exhibited in the fossil
record. Notably, with the development of nonlinear dynamics and complex network, it
is found that food web model presents a most suitable way to describe ecosystem.
More recent models based on network structure and various types of interaction have
produced convincing results not only for macroevolution but also for microevolution.
Microevolution that focuses on the influence of one species on others is
more important in protecting the environment, whereas macroevolution, which focuses on
species coevolution and periodic species extinction, is more interesting and
important for species diversity.

Many remarkable phenomena have been found in the fossil record but, until now no
satisfactory explanation has been presented. These phenomena include the Cambrian
explosion that has given rise to all presently existing animal phyla, the
periodicity of mass extinctions and characteristics of the evolution, such as
extinction rate, origination rate, percent extinction, standing diversity,
survivorship and lifetime distribution.

Recently, there have been attempts to study macroevolution using models equipped
with dynamics that operate at the level of individuals~\cite{791,792,793}.
Lotka-Volterra models are relatively successful in describing many aspects of
population dynamics. Coppex has introduced a simple two-trophic network to describe
an ecosystem with $N$ species of predators and one species of prey~\cite{F. Coppex}.
His numerical simulations show that there is a power-law distribution of intervals
between extinctions. Since the Lotka-Volterra model is a differential equation
model, it is hard to simulate the global-level processes with a large value of $N$.
Furthermore, his model does not describe the macroevolution following the beginning
of the Cambrian. Luz-Burgoa and cooperators have investigated the process of
sympatric speciation in a simple food web model~\cite{Luz-Burgoa}. They find that
sympatric speciation is obtained for the top species in both cases, and their
results suggest that the speciation velocity depends on how far up, in the food
chain, the focus population is feeding. In their later work a self-organized model
without the need of two different resources has been presented~\cite{Luz-Burgoa1}.
However, these models investigate only the cause of species speciation and cannot
explain the splitting process of a phylum.

B.F De Blasio and F.V. De Blasio have introduced competition interaction into the
computer model of the influence of ecospace colonization on adaptive radiation
designed by J. W. Valentine and T. D. Walker~\cite{valentine,AR}. This macroevolution
model presents a clear picture of the simultaneous appearance of so many phyla in
the Cambrian and the post-explosion evolution of the ecosystem. However, this model
cannot show an accurate picture of macroevolution. It lacks periodicity of extinction
and the results presented do not describe whether this model is consistent with
the characteristics of the real fossil record. Moreover, this competition rule seems
too simple to describe the intricate interaction between different species.

In fact, such interspecies competitive interactions play a key role in
macroevolution whether considered as the sum of predator-prey interaction,
host-parasite interaction or competitions between species using the same resource.
Competition also appears to be essential for species proliferation. Laboratory
bacterial experiments have suggested that species branching is promoted by
competition~\cite{AR12,AR13} and competition among higher taxonomical groups may
also play a major role in macroevolution~\cite{AR15}.

In this paper we present a macroevolution model also based on competitive
interactions. In contrast to the model in Ref.~\cite{AR}, our model is built on the
growth of network structures. From our simulation results of the properties, such as
degree distribution, clustering coefficients, and especially the modularity
dynamics, one can gain more insight into the processes of creating and splitting
phyla and the possible effects of mass extinction. Additionally, we observe the
periodicity of mass extinctions without any extraterrestrial causes (not mentioned
in the model of Ref.~\cite{AR}). This phenomenon is suggested by Raup and Sepkoski
in $1984$~\cite{periodicityExtinc}. Our model indicates that this phenomenon,
periodicity of mass extinctions, might be a natural consequence of macroevolution
and not the result of any extraterrestrial causes as predicted by the model of
Lipowski~\cite{add1,add2}. Furthermore, the simulation results of lifetime
distribution are very similar to the fossil record and the normal experiences of
human beings. Finally, in order to study the effects of mass extinctions, large
numbers of species were removed from network.

This article is organized as follows. We will explain how to build our model in Sec.
\ref{sec2}. In Sec. \ref{sec3}, simulation results are shown and discussed,
including network structure properties, lifetime distributions, extinction dynamics,
and perturbations of the model. In the last Section, we summarize our conclusions
and suggest some further extensions of our work.

\section{NETWORK MODEL\label{sec2}}

We take every species as a point in 2-dimensional morphospace without boundary and
let each dimension be independent. Each dimensional morphospace represents a
phenotypic character of the species such as shape, form, structure, and so on. We select
2-dimensional morphospace as a compromise between integrated description of species
and fast computation. Every species is characterized by a point in the morphospace.
Species that belong to the same phylum are similar in phenotypic character, and they
are close in morphospace. Phenotypic similarity between species implies that they
are likely to exploit the same resources~\cite{AR16}. However, even species that are
far from each other morphologically may compete for common resources, like the
competition for light among plants and water among animals. Long-range competition
is accounted for by allowing a finite tail in the possible competitive interaction
for morphologically distant species. The model dynamics are subsequently described.

(1) {\it Architecture of the network.} Every node in the network represents one
species and the weight of the connecting edge indicates the competition strength
between two species. The evolution of the network represents the evolution of all
species. We give the weight on the edges following a Gaussian-like form. It is
chosen from the interval $[0,1]$, weighted as a truncated Gaussian of mean $0.5$ and
the variance of the original, nontruncated Gaussian is also $0.5$. Once an
edge is built, it does not change. However, when a species becomes extinct, we
remove this node and all the edges connected to it from the network.

(2) {\it Initialization.} At the beginning, three species with the same phenotype
are connected as the initial network. Then we assign the weights of three edges.

(3) {\it Competition.} Since not all species compete with each other even if
they are very close in morphospace, the network model should not be a fully
interconnected graph. The competition strength is equal to the weight of the edges.
The extinction probability per step $p_{ext}$ of species $s_i$ depends on the total
weight of edges connected with it. So the extinction probability of one species is
defined by,

\begin{equation}
 p_{ext} (s_i)=\varepsilon\sum_{j}g_{ij},
\label{eq-1}
\end{equation}
where the $g_{ij}$ is the weight of the edge between species $s_i$ and $s_j$, and
the fraction of competition $\varepsilon$ is constant. If one species becomes
extinct, it will be erased from the network immediately and other species will no
longer experience competition from the extinct one. As the network develops, the
degree of the nodes increase and they will also feel more competition. The larger
$p_{ext}$, the more easily species $s_i$ becomes extinct. Therefore, the $p_{ext}$ is
hardly close to 1. If the $p_{ext}(s_i)$ is larger than 1, we remove this species from
the system.

(4) {\it Speciation.} In the short range, it appears to be approximately correct to
say that the whole world can support a certain number of species. Modern-day
ecological data on island biogeography support this view~\cite{Rosenzweig}. It is
reasonable to assume that new species are more difficult to evolve as the species
number approaches to this upper limit which is higher than that when the species is
rare. When a new species originates from an old species, the distance in the
morphospace between this new species and its parent for each dimension is
Gaussian-distributed with mean $0$ and the standard deviation $\sigma$. Here, the
$\sigma=0.1$ is the same for all simulations. New species is formed at each time step
with the speciation rate,

\begin{equation}
p_{sp}=\beta\left(1-\frac{N}{N'}\right), \label{eq-2}
\end{equation}
where $\beta$ is the fraction of the speciation rate, $N'$ is the maximal species
number in the world, and $N$ is the current species number or node number in the
network. In this work, the maximal species is set to $N'=5500$ for all simulations.

After a new species was born, it becomes a part of the network, but it is isolated from
other nodes. The probability $f(d_{ij})$ that the new node $i$ will be connected to
the existent node $j$ ($j=i$ is excluded) depends on the distance between them in
the morphospace.

\begin{equation}
f(d_{ij})= \left\{
             \begin{array}{ll}
               f_0, & d_{ij}<R; \label{eq-3a}\\
               f_\infty, & d_{ij}\geqslant R.
             \end{array}
           \right.
\end{equation}

Here the $d_{ij}$ is the distance between species $s_i$ and $s_j$. We set the radius
of short-range competition $R=2$ for all simulations, which was chosen larger than
the speciation range $\sigma$ so that speciation takes place within the range of the
short-range competition. The short-range competition factor $f_0$ should be much
larger than the long-range competition factor $f_\infty$.

In this model, the reason for the survival of species alive is not its intrinsic morphological
advantages but the morphological disparity of a species. In other words, the fitness
landscape is flat. In each time step, we chose one species randomly and decided
whether it would evolve into a new species based on the speciation rate. Then, we
randomly selected another species and decided whether it should be removed from the
network at that time step. This procedure is repeated millions of times in our
simulations.

\section{\label{sec3}SIMULATION RESULT AND DISCUSSION}

In this section we present the simulation results of our network model, such as
degree distribution, clustering coefficient and community structure. Moreover, to
study macroevolution we take the lifetime distribution, the extinction dynamics, and perturbation of network growth
into account in our simulations .

\subsection{\label{sec3a}Time Evolution of Network Structure}

\begin{figure}
\begin{center}
\includegraphics[width=0.5\textwidth]{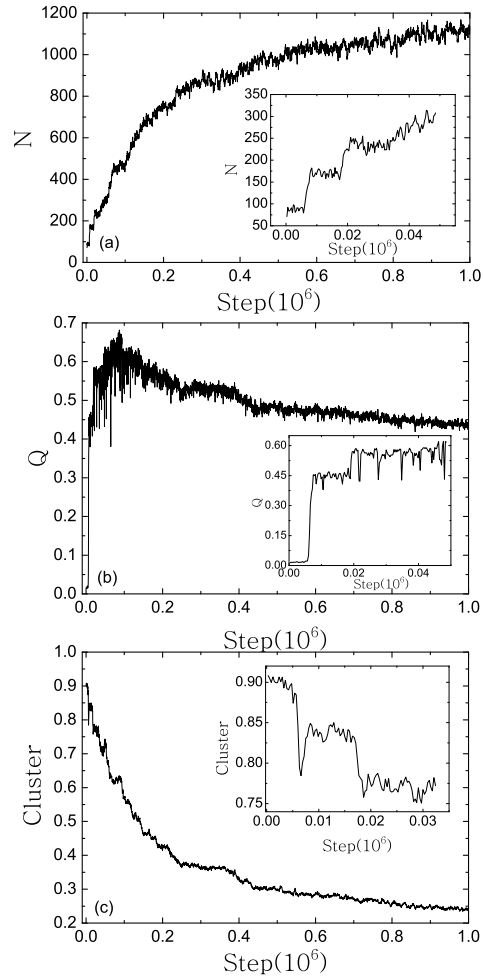}
\caption{Temporal evolution of the structural properties of the network with
parameters $\beta=0.6$, $\varepsilon=0.015$, $f_0=0.9$, and $f_\infty=0.03$. The
insets show the magnification of the first $40000$ steps. A step-like increase is
observed in the inset picture and corresponds to the split of the phyla.}
\label{fig1}
\end{center}
\end{figure}

\begin{figure*}[t]
\begin{center}
\includegraphics[width=1\textwidth]{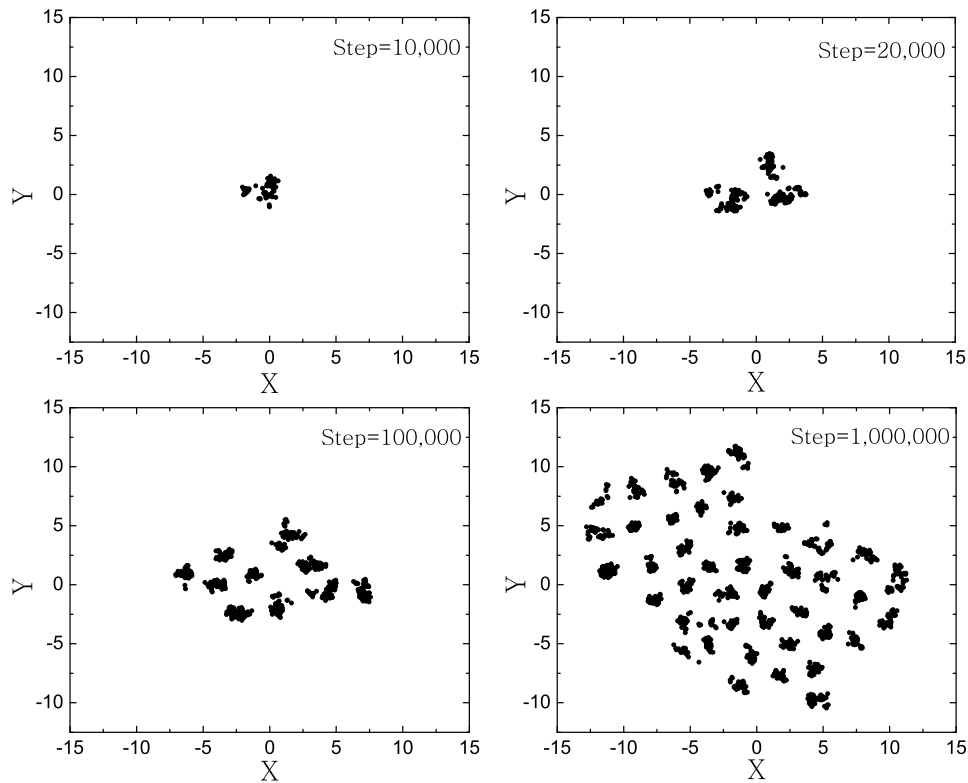}
\caption{Species distribution in morphospace at four different times. Top left,
$10,000$ time steps after the beginning of simulation; top right, after $20,000$
time steps; bottom left, after $100,000$ time steps; bottom right, after $1,000,000$
time steps. The figure shows the dynamics of phylum split and the phylum branching
process in morphospace. (More details about the splitting process can be seen at
supplemental materials: \url{http://www.youtube.com/watch?v=iJllf_tiaw0} and
\url{http://www.youtube.com/watch?v=87qqmMM1Dqc}) The parameters of simulation are
identical to those in Fig.~\ref{fig1}.} \label{fig2}
\end{center}
\end{figure*}

\begin{figure}
\begin{center}
\includegraphics[width=0.5\textwidth]{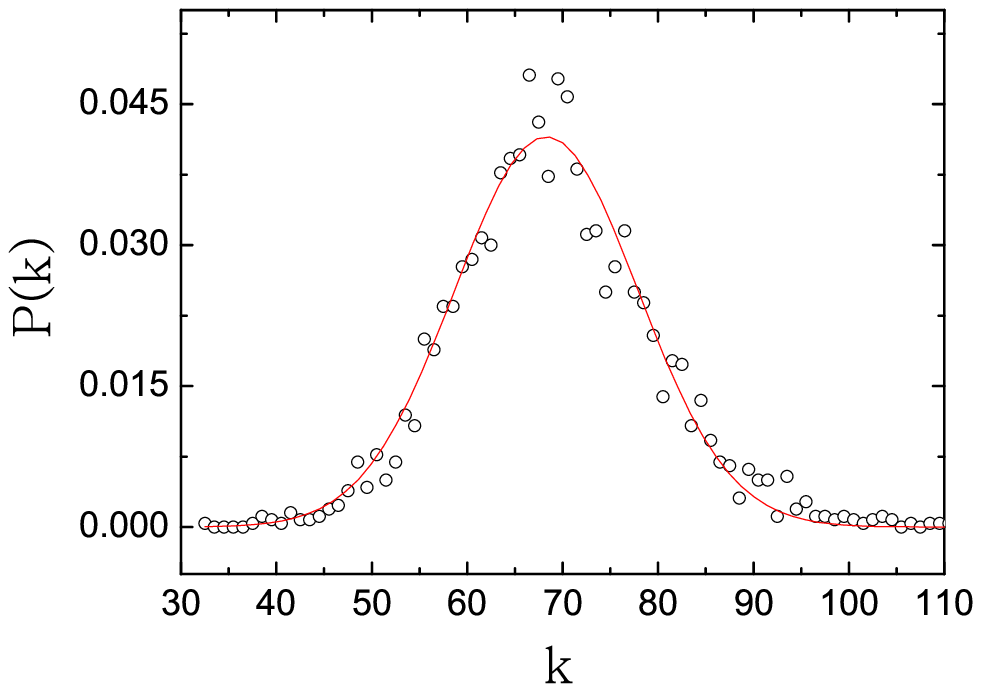}
\caption{(Color online) Degree distribution of our network model with saturation
node number $N\approx2600$ after adequate steps simulation. Parameter values are
$\beta=1.0$, $\varepsilon=0.015$, $f_0=0.5$, and $f_\infty=0.01$. The red line is a
Gaussian fitting result with the mean $68.29184$ and variance $9.61$.} \label{fig3}
\end{center}
\end{figure}

In general, two statistical properties, degree distribution and clustering
coefficient, are used to measure the structures of complex networks. Clustering
represents a common property of social network. Social network contains circles of
friends or acquaintances in which every member knows each other. This
tendency is quantified by the clustering coefficient. In our model, we use it to
present the circles of competition existing between three species. For every node $i$
in the network, we could get $C_i=\frac{E_i}{k_i(k_i-1)/2}$, where $k_i$ is the
degree of it and $E_i$ is the number of edges that existed between node $i$'s $k_i$
neighborhoods. The clustering coefficient of the whole network is the average of all
individual $C_i$.

To study phylum splitting, we calculate the community structure of networks.
Community structure is a mesoscopic description of networks. Newman and Girvan
\cite{newman04,girvan02} refers to the fact that nodes in many real networks appear
to group in subgraphs. Like this property in the real networks, all the species can
also be divided into different phyla. Therefore, the concept of community is used in our model
to observe the phylum splitting. In our model, there is only one taxonomic group, so
the community can be interpreted as genera, families, orders, classes, or any other
taxonomic group. The strength of the community structure can be quantified by the
modularity $Q$. Theoretically, the concept of modularity $Q$ cannot be used in
weighted networks. Even then, community structure is still observed in the
structural evolution in our simulations. The main reason for this observation is
that in our model the connections among the nodes play a major role in the
interspecies competition. Therefore we focus our attention on the structures and
ignore the differences between the edges. In addition, we used the algorithm to find
the community, as in Ref. \cite{algorithms}, which is based on an extremal
optimization of the value of modularity and is feasible for the accurate
identification of community structure in large complex networks.

It should be noted that our model is a network with dynamic node number.
Macroevolution is described by the process of network growth. In Fig.~\ref{fig1}, we
present how the species number, modularity and clustering coefficient evolve in the
simulation. The main character of our simulation is that the network gradually
stabilizes after an initial period of strong diversification. The inserted figure
shows that the increasing species number and decreasing modularity and clustering
coefficient are stepwise at the early phase. It should be noted that in the inserted
figure the discrete transitions for species number, clustering coefficient, and
modularity occur at the same time. These transitions denote the phylum splitting.
When the model is on the plateau stage, the system is in a temporary stable state.
The balance exists between the competition and the speciation rate. The process of
phylum splitting is interesting. When the species get together to form a stable
phylum, the diameter of every phylum on morphospace is about $R$ and the species in
the phylum interact with short-range competition. Because there is a small distance
between a new species and its parents, only the species on the fringe of the
phylum contribute to the diffusion. The diameter of the phylum in morphospace
increases. As the diameter of the phylum is larger than $R$, the central species of
one phylum which is under strong competition from the fringe species are easily to
extinct. The shrinking of central species facilitates the diffusion of fringe
species. Diffusion and shrinking are not asymmetrical. Therefore, one phylum splits
from the center and two parts separate gradually to form new phyla. Because the
distance between two new phyla is larger than $R$, species will meet with more
long-range competition from another phylum. The average number of species in the
phylum will be smaller than that before splitting. Then the more phyla, the fewer
species in every phylum. After the rapid increase in the initial phase, species
proliferation decelerates to a constant value with little fluctuation. The
modularity and clustering coefficient also decrease gradually while the species
number increases slowly. This suggests that the number of community is increasing.
This process can also be observed in Fig.~\ref{fig2}.

The beginning increase of stepwise modularity can be interpreted by the definition
given by Newman, $Q = \sum_{\gamma} (e_{\gamma\gamma} - a_\gamma^2)$, where
$e_{\gamma\gamma}$ is the fraction of edges that connect two nodes inside the
community $\gamma$, $a_\gamma$ the fraction of links that have one or both nodes
inside of the community $\gamma$. In our model, the definition can be written in the
following form:

\begin{equation}
Q =\sum_\gamma\left(\frac{2z_\gamma^{in}}{2m}-\frac{2z_\gamma^{in}+z_\gamma^{ex}}
{\left(2m\right)^2}\right), \label{eq-4}
\end{equation}
where $m=\frac{1}{2}\sum_\gamma(2z_\gamma^{in}+z_\gamma^{ex})$ is the number of
edges in the network, $z_\gamma^{in}$ is the number of edges which connect two
nodes inside the community, and $z_\gamma^{ex}$ is the number of intercommunity
edges. So only the network has many phyla and a large number of species in every
phylum have the large $Q$. Considering that $f_0=0.9$ is much larger than
$f_\infty=0.03$, we assume $z_\gamma^{in} \gg z_\gamma^{ex}$ and
$z_\gamma^{in}=z_\delta^{in}$. We get

\begin{equation}
Q(n)\approx 1-1/n, \label{eq-5}
\end{equation}
where $n$ is the community number in the network.

After the rapid increase in the initial phase, species proliferation decelerates to
a constant value with little fluctuation. During this stage, the number of phyla is
still increasing slowly. This phenomenon can be observed in Fig.~\ref{fig2}(d).
However, as discussed above, with increasing phyla in morphospace, the number of
species in every phylum will decrease gradually. When the increase of $\gamma$
cannot compensate for the decrease of $z_\gamma^{in}$ and increase of
$z_\gamma^{ex}$, the assumption $z_\gamma^{in} \gg z_\gamma^{ex}$ cannot be
fulfilled anymore. Therefore modularity $Q$ will descend gradually. That is the
reason for the decreasing modularity after the initial stage.

In Fig. ~\ref{fig2}, we present simulation results of the species distribution in
morphospace at different times. Every point in the morphospace denotes a species.
Obviously, species congregate around to form phylum. The figure on the bottom right
is the species distribution of $1,000,000$ steps when the system has been on the
stable stage and, once in stable stage, the number of phylum will not increase
monotonously with the time, but every phylum's position move slowly.

In order to understand the properties of the network model for biology, one must
have some knowledge about the degree distribution $P(k)$, which is the probability
that a node has degree $k$. Note that the number of vertices of this network model
is dynamic. To study $P(k)$, it should be assigned large $\beta$ and other
parameters as small to generate a network with a large number of nodes. In Fig. \ref{fig3},
we present the degree distribution of the network with $N \approx 2600$ at the final
saturation state. It was found that P(k) is very close to a Gaussian distribution
with mean $k\approx68$. This distribution can be understood by a two-level random
network. One is caused by the short-range competition, which has a few nodes in one
community with connecting probability $f_0$. The other is caused by the long-range
competition, which has many nodes whose number is almost equal to the whole nodes in
the network with a connecting probability $f_\infty$. We know that in random
networks, the modularity is very small and even close to 0. Therefore, it is
difficult to categorize our network model as a type of network such as random,
scale-free, or small-world.

\subsection{\label{sec3b}Lifetime Distribution}

\begin{figure}
\begin{center}
\includegraphics[width=0.5\textwidth]{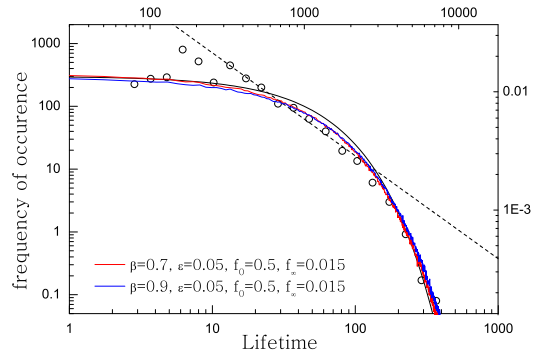}
\caption{(Color online) Histogram of lifetimes of marine genera during the
Phanerozoic (black circles). The black dash line is the best power-law fit to the
points between 10 and 100 My and the black solid line is the best exponential fit to
all the points ~\cite{lifetimehistory,newman99}. The red and blue solid lines are
our simulation results from the network growth model using two different parameters.
} \label{fig4}
\end{center}
\end{figure}

One of the properties frequently studied in macroevolution models is lifetime
distribution of species. Newman and Sibani mathematically derived a number of
relations between the normal quantities (extinction rates, diversity, lifetime
distribution, etc.) and showed how these different trends are inter-related
\cite{newman99}. Since the lifetime distribution is easy to calculate, it becomes a
canonical simulation result in most models of ecosystems.

Using simulations based on various models, many scholars have measured the lifetimes
of competing species and suggested that their distribution is well approximated by a
power-law form \cite{newman99}. Similar estimations demonstrate that this
distribution is equivalent to the power-law distribution of genus lifetimes, since
the longer lived genera give rise on average to larger numbers of species
\cite{adami95}. However, the real fossil records are fit equally well by an
exponential form \cite{newman99}.

In Fig. \ref{fig4} we show a histogram of species lifetime distributions for the
network growth model with two different parameters $\beta$. As the plot shows, the
distribution closely follows an exponential law. Note that two simulation results of
the lifetime distribution of species for different $\beta$ are very similar. We
conjecture that the lifetime distribution of species is not under the influence of
the fraction of the speciation rate $\beta$, but depends on the ratio between $f_0$
and $f_\infty$ (or the ratio between competition within a phylum and competition for
resources). Moreover, this observation emphasizes that the competitive interactions
play a key role in ecological dynamics.

\subsection{\label{sec3c}Extinction Dynamics}

\begin{figure}
\begin{center}
\includegraphics[width=0.6\textwidth]{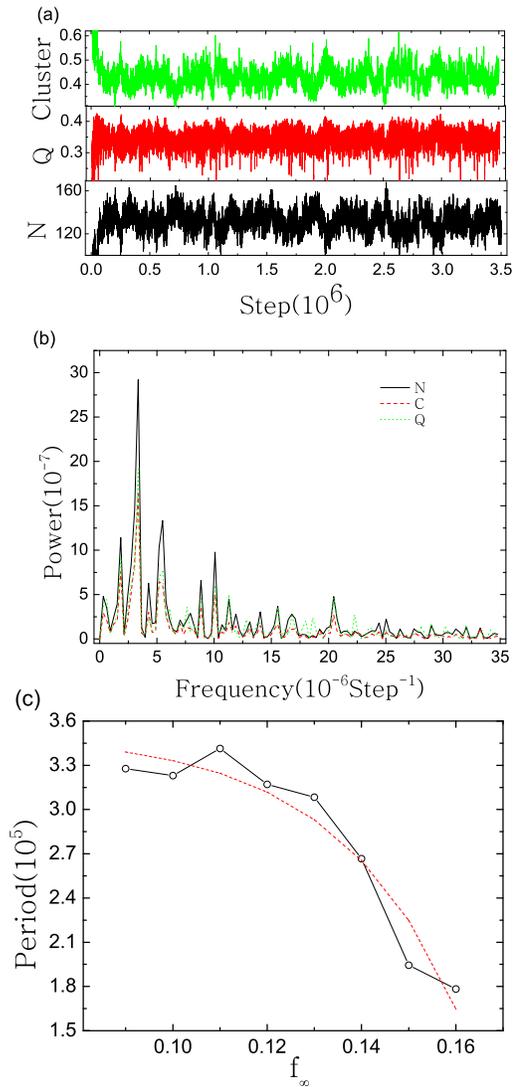}
\caption{(Color online) Periodicity of mass extinctions without an extraterrestrial
cause. Parameter values are $\beta=0.9$, $\varepsilon=0.05$, $f_0=0.9$, and
$f_\infty=0.14$. (a) Time series of the clustering coefficient, modularity $Q$, and
species number $N$. (b) The Power Spectral Density of the data in (a). (c) The
maximal periodicity of oscillation $\tau$ from the time series of species number $N$
as a function of $f_\infty$. The other parameters are the same as in (a).}
\label{fig5}
\end{center}
\end{figure}

Of the estimated one to four billion species which have existed on the Earth since
life first appeared here, less than 50 million are still alive today.
Paleontological data, which show broad distributions of the extinct events in the
Earth's history, suggest the existence of strong correlations between
extinctions~\cite{9908002}. Normally, the majority of researchers prefer the
alternative explanation that the extinctions appear because of external stresses
imposed on the ecosystems by the environment. Recently, the occurrence of
extinctions in the absence of periodic external perturbation was suggested by
Lipowski in a lattice model~\cite{add1,add2}.

Fig. \ref{fig5} shows the temporal evolution of $N$, $Q$, and the clustering
coefficient at the saturation state with parameters $\beta=0.9$, $\varepsilon=0.05$,
$f_0=0.9$, and $f_\infty=0.14$. Note that we omit the initial phase of the network
growth in Fig. \ref{fig5}(a). The Power Spectral Density of the data is presented in
Fig. \ref{fig5}(b). In order to compare the periodicity of $N$, $Q$ and the
clustering coefficient, we subtract their means and rescale these data to make them
have the same amplitudes before the Fourier transform. The peak in Fig.
\ref{fig5}(b) shows that the species number $N$ and the modularity $Q$ and the
clustering coefficient of the network have the same periodicity and power spectral
density. Three curves have the same peak at frequency $3.35693(\times
10^{-6}Step^{-1})$. The maximal periodicity of oscillations $\tau$ for various
$f_\infty$ is shown in Fig. \ref{fig5}(c). The red dashed line is a fitted result
$\tau=351410-375.074\mathtt{exp}(38.81 \times f_\infty)$. Therefore this oscillation
indicates that the extinction is on the phylum level. The oscillation of the network
structure parameters reflects the number and scale changing of phylum. Phylum which
bears intense competition will decrease its scale or even this phylum will go
extinct. In most instances, large numbers of phyla do not routinely go extinct, and
they may belong to extant groups rather than constituting distinct phyla.

It should be noted that the species number in Fig. \ref{fig5}(a) is very small, $N
\approx 140$, due to the larger long-range competition $f_\infty =0.14$. In fact, by
enlarging the simulation results of the saturation phase in Fig. \ref{fig1} with
$f_\infty =0.03$, it was found that similar oscillation behaviors take place in the
large networks. However, the oscillation in Fig. \ref{fig1}  appears to be
small-amplitude fluctuations.

It is clear from Fig. \ref{fig5}(c) that the periodicity of oscillation is a
function of $f_\infty$. Moreover, our extensive simulation results demonstrate that
there is no obvious correlation between the periodicity and other parameters. When
the system is in stable stage, there is a balance between the extinction rate and the
speciation rate. The extinction rate is caused by the short-range competition coming
from the species in the same phylum and long-range competition coming from the
species in another phylum. The process of phylum splitting increases the species
number and phylum number. Then the balance is disturbed. The species in the phylum
will bear more competition from other phyla. Therefore the mass extinction happens.
If $f_\infty=0$, new phylum would not be affected by other phyla. The mass
extinction would be unconspicuous either. Therefore we conjecture that the
long-range competition for public resources induces the oscillation of the species
number. However, smaller $f_\infty$ or a scenario with abundant public resources
would not only increase the system size $N$ and the phylum number, but also decrease
the extinction of species and phyla. As a result, only a small fluctuation of
species and phyla exists.

\subsection{\label{sec3d}Perturbing the Model}

\begin{figure}
\begin{center}
\includegraphics[width=0.5\textwidth]{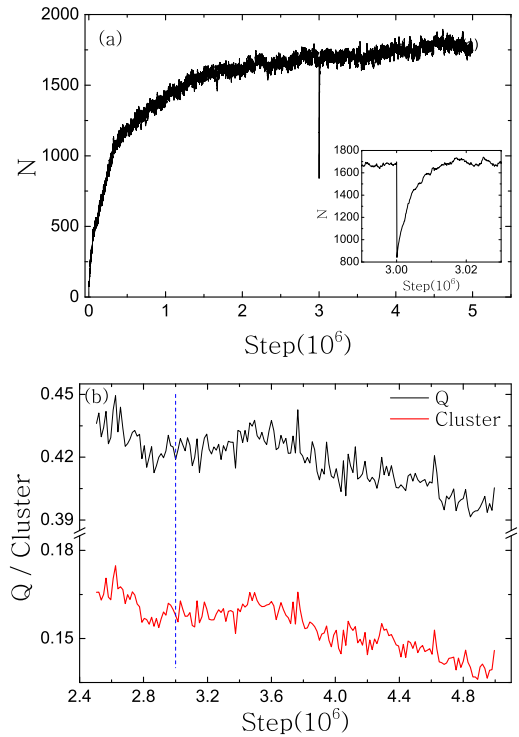}
\caption{(Color online) The network growth with a random extinction at $3 \times
10^6$ steps. (a) Time series of the species number. The inset shows a magnification
of the extinction region. (b) The modularity and clustering coefficient in the extinction region.
The dotted line is the time of extinction. The parameters are $\beta=0.5$,
$\varepsilon=0.015$, $f_0=0.7$, and $f_\infty=0.015$ .} \label{fig6}
\end{center}
\end{figure}

\begin{figure}
\begin{center}
\includegraphics[width=0.5\textwidth]{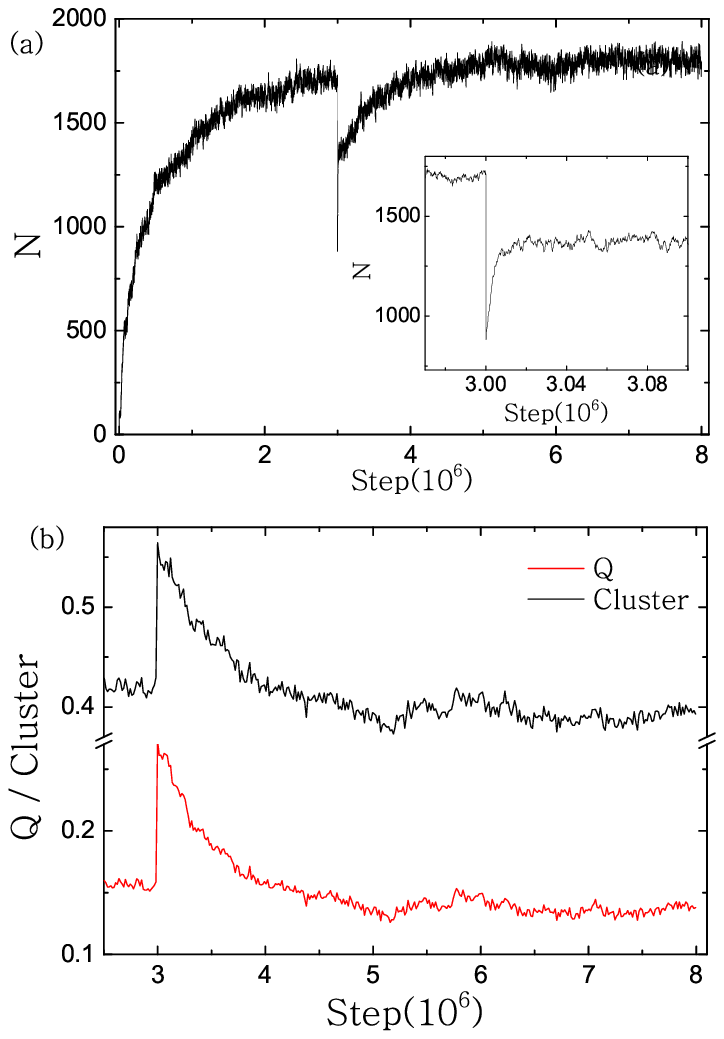}
\caption{(Color online) The network growth with a random community extinction at $3
\times 10^6$ steps. (a) The time evolution of the species number. The inset shows a
magnification of the extinction region. (b) The modularity and cluster dynamics in
the extinction region. The dotted line is the time of extinction. The parameters are
identical to those used in Fig.~\ref{fig6}.} \label{fig7}
\end{center}
\end{figure}

In order to study the effect of mass extinctions on the macroevolution with the
network growth approach, a large number of species were removed from the network at
the saturation state. Such a scenario is designed to mimic the influence of an
external factor, such as the change of climate, the impacts of comets or meteorites,
etc. We discuss the following two cases.

Case I: {\it Random perturbation on species}. In this case, half of the species were
removed randomly. Such a situation is appropriate for biosphere in which the species
become extinct because of 'bad luck'. Clearly, the dynamics of the network should be
reduced to the saturation state quickly since each species has the same probability
of extinction.

Case II: {\it Random perturbation on community}. In this case, the artificial
perturbations act on the community. We remove half of the community in the network
randomly. The motivation for this case is based on the idea that some phyla have
'bad genes' and cannot adapt to the sudden change of environment.

Figs. \ref{fig6}-\ref{fig7} show the simulation results of the time series of the
species number and the network structure for both cases. Obviously, even though both
scenarios of mass extinction act on different hierarchic levels, the species in the
world will eventually recover and the response to an extinction will include a rapid
expansion of species diversity. However, there is a noticeable difference, which can
be seen in Fig. \ref{fig6}-\ref{fig7}, in the detailed recovery process.

In case I, after removing nodes in the network, the species number quickly returned
to the pre-perturbation level (see Fig. \ref{fig6}(a)). Fig. \ref{fig6}(b) depicts
the response of network structure to extinction. There is not any significant change
of the modularity and clustering coefficient and both are still in the normal
fluctuation range.

In contrast to case I, the recovery in case II is slow and difficult. Many more
steps are necessary for the species number to reach the saturation stage (see Fig.
\ref{fig7} (a)). However, note that there is also a sharply rising stage (as shown
in the inset of Fig. \ref{fig7}(a)). The initial stage of the response to the
extinction, when compared with the inset of Fig. \ref{fig6}(a), is similar for both
cases. In Fig. \ref{fig7}(b), we present the responses of the modularity and
clustering coefficient to this extinction. These curves represent a sudden increase
preceding the decrease of the diversity at the phylum level.

As demonstrated by the arguments given above, it is clear that the response of the
biological evolution for the two scenarios of mass extinction is completely
different. Although both perturbations erase almost the same ratio of the species,
in Case I the perturbation does not break down the structure of the network.
In Case II, however, a number of whole communities are removed, or some phyla become
extinct. Therefore the network structure is destroyed and the perturbation of Case
II is more deleterious.

The statement above implies that forming phyla is a way for species to avoid or
replace the short-range competition and increase the species number. Therefore it is
much easier for species to conquer short-range competition for speciation than to
develop long-range competition. In fact, in Case I, removing nodes in the network
decreases both long- and short-range competition. The new species do not need to
develop new phyla. This is the explaination of the fast recovery after disturbance. In
contrast, Case II decreases the long-range competition. The new species develop in
their phyla because the absence of long-range competition decreases the short-range
competition below saturation and the small $\sigma$ make new species remain in their
phyla. After this stage, when every phylum is full of species and reaches
saturation, the new species develop new phyla.  These new phyla are booming and
force old phyla until they have the same scale. Thus, the behaviors of the
modularity and cluster are different in two scenarios following disturbances, and
the species number increases slowly to the saturation level after the initial
sharply rising stage.

\section{Conclusion\label{sec4}}

In this paper, we present a simple network growth model with competitive
interactions to approximate the biological evolution. This model is characterized by
four tunable parameters: the fraction of competition $\varepsilon$, the fraction of
speciation rate $\beta$, the short-range competition $f_0$, and the long-range
competition $f_\infty$. We start with a few species, and then let our model grow in
diversity and complexity until it reaches the saturation state. It should be noted
that the limitation $N'$ plays an important role in the evolution. Without this
parameter, periodical extinction cannot be displayed. These simulation results are
not presented in this paper. Our model can be established on the higher dimensions,
but it is difficult to show the community beyond the two-dimensional case. In a
higher dimension case, a larger saturation species number on the stable state is
formed. These results are not shown in this paper.

Based on the simulation results of our network growth model, we present several
different aspects of biological evolution, such as the species number, phyla,
clustering coefficient, and the lifetime distribution. We also observe the periodic
mass extinction. Finally, the network model is perturbed in two different ways:
random perturbation species and random perturbation on community. The effects of
these different extinction scenarios on the two taxonomic hierarchic levels, species
and phyla, may be useful to interpret of the causes of historical mass
extinction.

Our model is aimed at emphasizing the importance of competition and network
structure in the biological evolution. Competition not only decides the species
number in the whole system, but also makes the species split to different phyla. As
there are two kinds of different competitions, the responses of the biological
evolution for two mass extinctions are completely different. Competition coupled with
network structure drives the number of species to fluctuate periodically.

Although many of the properties shown in our model are similar to the fossil records
especially the lifetime distribution, it is difficult to compare the model with the
actual history of phyla diversities and real world. Since this model is a
macroevolution model, it cannot include all the properties and details of biology.

Further extensions of our model offer interesting opportunities. For example, one
can change the form of the competing function or differently implement the
interaction among species. This could complicate the simulation tremendously, and we
believe that the results would have the similar qualitative properties. In the real
world, the climate switches between the ice age and warmth within a period of about
$10,000$ years\cite{iceAge}. Comparably, our model could also use the periodic change
parameters instead of the constant form. Such a modification would likely result in
other thought-provoking phenomena. However, this modification is not currently
reasonable because we cannot assess whether the changing climate is strong enough to
affect the parameters in our model or whether a relationship exists between the
climate and the model.

\ack{ We thank the anonymous referees for constructive suggestions. This work was supported in part by the National Natural Science Foundation of China under Grant No. $10305005$ and by the Special Fund for Doctor Programs at Lanzhou University.}

\end{document}